\documentclass[prb,twocolumn,showpacs,superscriptaddress]{revtex4}
\usepackage[english]{babel}
\usepackage{amsmath,amssymb}
\usepackage{graphicx}
\begin{document}
\title{Electron Spin Resonance of SrCu$_{2}$(BO$_{3}$)$_{2}$ at High Magnetic Field}
\author{S. \surname{El Shawish}}
\affiliation{J. Stefan Institute, SI-1000 Ljubljana, Slovenia}
\author{J. \surname{Bon\v ca}}
\affiliation{J. Stefan Institute, SI-1000 Ljubljana, Slovenia}
\affiliation{Faculty of Mathematics and Physics, University of Ljubljana, SI-1000 Ljubljana, Slovenia}
\author{C.~D. \surname{Batista}}
\affiliation{Theoretical Division, Los Alamos National Laboratory, Los Alamos, NM 87545, USA}
\author{I. \surname{Sega}}
\affiliation{J. Stefan Institute, SI-1000 Ljubljana, Slovenia}

\date{\today}
\begin{abstract}
We calculate the electron spin resonance (ESR) spectra of the
quasi-two-dimensional dimer spin liquid SrCu$_{2}$(BO$_{3}$)$_{2}$ as
a function of magnetic field $B$. Using the standard Lanczos method,
we solve a Shastry-Sutherland Hamiltonian with additional
Dzyaloshinsky-Moriya (DM) terms which are crucial to explain different
qualitative aspects of the ESR spectra. In particular, a
nearest-neighbor DM interaction with a non-zero $D_z$ component is
required to explain the low frequency ESR lines for $B \|c$. This
suggests that crystal symmetry is lowered at low temperatures due to a
structural phase transition.
\end{abstract}
\maketitle

\section{Introduction}
The spin liquids are states of matter that occur when quantum
fluctuations are strong enough to avoid any type of magnetic ordering.
This leads in general to a non-degenerate ground state and a finite
gap for the spectrum of excitations. SrCu$_{2}$(BO$_{3}$)$_{2}$ is a
quasi-two-dimensional spin system with a singlet dimer ground state
\cite{smith91}. This compound is at p[resent the only known realization of the
Shastry-Sutherland model \cite{shastry81}. In this model the effect of
the quantum fluctuations is amplified by the geometrical frustration
of the spin lattice. The low energy excitations of the ground state
are local triplets whose ``kinetic energy'' is small compared to the
repulsive triplet-triplet interaction. The application of a strong
magnetic field induces a quantum phase transition in which the
dimerized ground state starts to be populated with triplets. The
magnetic field plays the role of a chemical potential for the triplet
quasi-particles. In this scenario it is possible to study the
competition between the ``gaseous'' triplet phase and the
crystallization of triplets (``solid phase''). The crystallization of
triplets gives rise to magnetization plateaus that have been observed
in SrCu$_{2}$(BO$_{3}$)$_{2}$ \cite{kageyama99,onizuka00}.

The anisotropic spin interactions are in general weak but they can
have a strong effect on a highly frustrated system. In particular, as
it was already established in
Refs.~\cite{cepas01,cepas02,jorge03,zorko00,zorko01}, the inclusion of
the nearest neighbor (nn) and next-nearest neighbor (nnn)
Dzyaloshinsky-Moriya (DM) interactions is required to explain some
qualitative features of the specific heat and electron spin resonance
(ESR) experiments in SrCu$_{2}$(BO$_{3}$)$_{2}$. However, as it was
noted recently by C\'epas {\it et al.} \cite{cepas01,cepas02}, a
lattice symmetry (reflection in the mirror plane containing the
$c$-axis and one dimer followed by a $\pi$ rotation arround the dimer bond)
leads to a zero amplitude for the observed single-triplet ESR transitions for
$B\|c$. In addition, a level anti-crossing between the ground state
and the lowest triplet excitation is observed for $B\sim 20$~T. This
level anti-crossing implies some mixing between two states with
different magnetization $M_z$ along the tetragonal $c$-axis, something
that cannot be explained within the $U(1)$ invariant models (invariant
under rotations around the $c$-axis) for which $M_z$ is a good quantum
number.

Recent experiments \cite{nojiri03} revealed additional quantitative
and qualitative aspects of the ESR transitions. Besides the two
non-degenerate one-triplet excitations, various types of multiple-triplet
bound states forming singlets, triplets and quintuplets were 
identified. These measurements opened the possibility for a direct
comparison between the observed family of magnetic excitations and the
theoretical predictions based on the spin model proposed for
SrCu$_{2}$(BO$_{3}$)$_{2}$ \cite{miyahara03}. In addition, as it is
shown in the present paper, they provide indirect information about
the crystal symmetry at low temperatures and the role of the
spin-lattice coupling as a function of the applied magnetic field $B$.

The considerable amplitude of the ESR absorption lines that according
to the crystal symmetry are not expected to be observed poses a
challenge for finding an adequate explanation.  C\'epas {\it et al.}
\cite{cepas02} proposed a mechanism based on a dynamically generated DM
interaction induced by the spin-phonon coupling.  However, they did not
provide any comparison between a calculated ESR spectrum based on this
mechanism and their experimental observation.  In this paper we
suggest that the crystal symmetry is lowered due to a structural phase
transition that occurs in a low temperature region that has not yet been
explored with X-rays. As a consequence, a non-zero $c$ component
of the nearest neighbor DM vector appears. In a previous paper,
\cite{jorge03} we showed that this component is required to reproduce
the measured specific heat at low temperatures and high magnetic
fields. Here, we show that the same component also explains the
observed single-triplet ESR lines as well as other qualitative aspects
of the ESR spectra as a function of $B \| c$ and $B \| a$.

\section{ Model Hamiltonian}

To describe the present system, we consider the following Heisenberg
Hamiltonian in a magnetic field on a Shastry-Sutherland lattice
\cite{shastry81}:
\begin{eqnarray}
\nonumber
H_s &=& J \sum_{\langle {\bf i, j} \rangle}  {\bf S_{i} \cdot S_{j}}
+ J' \sum_{\langle {\bf i, j} \rangle'} {\bf S_{i} \cdot S_{j}}
+ g\mu_B\sum_{\bf i} {\bf B}\cdot {\bf S_i}\\
&+&\sum_{\langle {\bf i \rightarrow j} \rangle}  {\bf D} \cdot
({\bf S_{i}\times S_{j}})
+ \sum_{\langle {\bf i \rightarrow j} \rangle'} {\bf D'} \cdot
({\bf S_{i} \times S_{j}}).
\label{Hamil}
\end{eqnarray}
Here, $\langle {i,j} \rangle$ and $\langle {i,j} \rangle'$ indicate
that ${\bf i}$ and ${\bf j}$ are nn and nnn, respectively. In addition
to the Shastry-Sutherland Hamiltonian, $H_s$ includes DM interactions
to $nn$ and $nnn$. The corresponding DM vectors are ${\bf D}$ and
${\bf D'}$, respectively.  The arrows indicate that bonds have a
particular orientation as described in Ref. \cite{jorge03}. The
quantization axis $\hat{\bf z}$ is parallel to the $c$-axis and
$\hat{\bf x}$ to the $a$-axis. The $nnn$ DM interaction
has already been considered in previous papers \cite{miyahara03} to
explain the splitting between the two single-triplet excitations
observed with ESR \cite{cepas01,nojiri03}, far infra-red \cite{room00}
and inelastic neutron scattering measurements \cite{kageyama00}. The
value of the DM interaction obtained from this splitting is: ${\bf
D'}=2.1~{\rm K}~{\hat{\bf z}}$. According to the crystal symmetry
{\cite{sparta01}, only the $xy$ component of ${\bf D}$ is non-zero and
perpendicular to the corresponding dimer. However, as it is explained
below, a non-zero $z$ component of ${\bf D}$ is required to explain
the specific heat and the ESR data for finite magnetic fields $B$.

For $D_z=0$ and in two dimensions the relevant space group of $H_s$ in
Eq.~(\ref{Hamil}) is $\rm p4gm$, with a point group $\rm 4mm$ at ${\bf q}=0$
point in the Brillouin zone. However, the
a-b plane in SrCu$_{2}$(BO$_{3}$)$_{2}$ containing dimers is  slightly
buckled and the (three dimensional) space group $\rm P{\bar 4}2_im$ is more
appropriate. Namely, the associated point group $\rm {\bar 4}2m$ includes
the roto-inversion symmetry $\rm IC_{4}^\pm$ (rotation through $\pm 
90^{\rm o}$ arround the c-axis followed by inversion $\rm I$) which allows
for a different orientation of the DM interaction with respect to the point
group $\rm 4mm$, and thus results in lowering the ground state energy. In
zero magnetic field $H_s$ is moreover time-reversal invariant so that the point
group at ${\bf q}=0$  should be enlarged to ${\rm G_{M}={\bar 4}2m}\times
\{{\rm E},\Theta\}$, where $\Theta$ is the time-reversal operator and $\rm E$
is the identity operator.

\section{Energy Spectrum}

Numerical calculations were done using the standard Lanczos technique
at zero temperature ($T=0$) on a tilted square cluster of $20$ sites.
We start by analyzing the full energy spectrum for ${\bf q}=0$.
Figs.~1a and 1b show the calculated energy spectra as a function of
$B\|c$ and $B\|a$, respectively. The vertical axis is in units of
frequency to facilitate the comparison with the ESR experimental
results by Nojiri {\it et al.} \cite{nojiri03}. With the exception of
$D^\prime_z$, the parameters of the model, $J=74$~K, $J^\prime=0.62J$,
${\bf D}=(2.2~{\rm K},2.2~{\rm K},3.7~{\rm K})$, and ${\bf
D^\prime}=(0,0,2.2~{\rm K})$, are the same as the ones used to fit the
specific heat data for different values of the applied magnetic field
\cite{jorge03}.

In both cases, there is a clear distinction between states that belong
to the continuum in the thermodynamic limit and those that will be
called ``localized'' states. The bottom edge of the continuum appears
around 1100~GHz above the ground state for $B=0$.  With increasing $B$
this edge drops to $\sim 450$~GHz at $B\sim 20$~T and then saturates
as a function of $B$. Since the DM terms are small compared to $J$ and
$J'$, the localized states can be classified according to their
approximate total spin quantum numbers, $S$ and $S_z$, in the regime
$B<20$~T. For $B<6$~T there are at least four singlet and two triplet
states split due to finite $B$ and $D_z^\prime$.
\begin{figure}[htb]
\includegraphics[angle=0,width=7.2cm,scale=1.0]{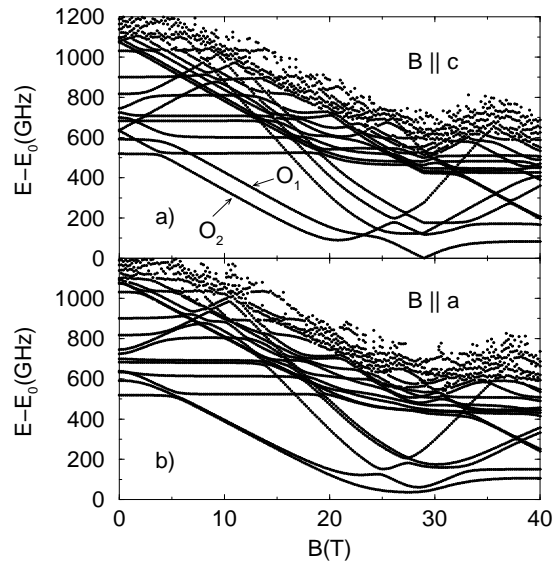}
\vspace{0.4cm}
\caption{Energy spectrum relative to the ground state energy $E_0$ of
the Hamiltonian $H_s$ (Eq.~(\ref{Hamil})) calculated on a $20$-site
cluster with periodic boundary conditions for a) $B\|c$ and b)
$B\|a$. The parameters of the model are: $J=74~{\rm K}$,
$J^\prime=0.62J$, ${\bf D}=(2.2~{\rm K},2.2~{\rm K},3.7~{\rm K})$ and
${\bf D^\prime}=(0,0,2.2~{\rm K})$.}
\label{fig1}
\end{figure}

As it is expected from the Zeeman interaction, for $B\|c$
(Fig.~\ref{fig1}a) the energy of the two $S_z=-1$ triplet states
($O_2$ and $O_1$ in the notation of Ref.~\cite{nojiri03}) decreases
linearly in the applied field. These two states are separated by an
energy $\delta \sim 90$~GHz and their position agrees well with
experimental ESR lines observed in Refs.~\cite{cepas01,nojiri03}.
Around $B_c=20$~T, the lower triplet state gets mixed with the singlet
ground state producing a level anti-crossing that is a consequence of
the finite values of $D_x$ and $D_y$. For $D_z=0$, only the lowest
triplet state $O_2$ mixes with the singlet ground state. This is
because the $O_1$ triplet state and the singlet ground state belong to
different representations of the point group $\rm G_M$.  
More precisely, a finite magnetic field breaks the $\rm \Theta$-invariance
and reduces the symmetry from the full point group $\rm G_M$ to the (magnetic)
subgroup $\rm {\bar 4}{\underline 2}{\underline m}$ \cite{hammer}. The
ground state and $O_2$ transform according to the               
irreducible representation $\rm A_1$ of $\rm {\bar 4}{\underline
2}{\underline m}$, while $O_1$ transforms according to $\rm B_2$.
The main difference between these representations is in the fact that a
roto-inversion with respect to the c-axis gives a sign $+1$ for the ground
state and $O_2$, while it gives $-1$ for $O_1$. 

The hybridization which is induced by a finite value of $D_z=3.7~{\rm
K}$ is too small to be observed with the ESR experiment in the absence
of $D_x$ and $D_y$terms. These findings are in agreement with the
experimental data \cite{cepas01,nojiri03}. For $B\|a$ the two triplets
are nearly degenerate except above $B=20$~T. Again, in agreement with
the experimental data \cite{cepas01,nojiri03}, the two triplet states
split around $B>20$~T because of the different hybridization between
each of them and the ground state. Note that the effect of this
hybridization becomes significant when the energy difference between
the triplets and the singlet ground state becomes comparable to DM
interaction.

The agreement with the experiment extends even further. The level
anti-crossing of triplet and singlet states around $646$~GHz and
860~GHz for $B\|a$ is also reproduced (see Figs.~4b and 7 in
Ref.~\cite{nojiri03}).  Fig.~\ref{fig1}b also shows a weak level
anti-crossing of the $O_1$ $S_z=-1$ triplet state with the first
singlet excitation located around 520~GHz. This effect seems to be too
small to be experimentally observable. However, a much stronger level
anti-crossing of the $O_2$ triplet with a singlet located around and
600~GHz is observed at $B\sim 2.5$~T. Another strong level
anti-crossing of the $O_1$ triplet with a singlet bound state is
observed near $B\sim 4$~T and 800~GHz. Although the values of the
magnetic fields of these anti-crossings are in good agreement with the
experiment, the calculated frequencies deviate from the observed
values. A comparison of the calculated ESR spectra for different
cluster sizes suggests that this deviation is due to finite-size
effects. For $B\|c$, the $O_1$ triplet produces strong level
anti-crossings for $B\sim 4.5$~T, $\nu=600$~GHz and $B\sim 2.5$~T,
$\nu=800$~GHz. Experimentally, only the later crossing is clearly
visible in Fig.4a of Ref.~\cite{nojiri03}.

\section{Spin Structure Factor}
To make account of the frequency and the intensity of the ESR lines we
need to compute the dynamical spin structure factor for ${\bf q}=0$,
\begin{equation}
S^\mu(\omega) = {\frac{1}{\pi}}{\rm Re}\int_0^\infty\!\!\! dt\ 
e^{i\omega t}\langle S^\mu(t)S^\mu(0)\rangle;\ \mu=x,y\  {\rm or}\  z,
\label{som}
\end{equation}
in the direction perpendicular to the applied magnetic field. The
method that we used to compute $S^\mu(\omega)$ is described in
Refs.~\cite{gagliano,jprev}. Fig.~\ref{fig2} shows the computed ESR
spectrum as a function of frequency $\nu=\omega/2\pi$ and the external
magnetic field $B$ along the $c$- and the $a$-axis. We use this
frequency-field type of  diagram to directly compare with the
experimental data obtained by Nojiri {\it et al.} \cite{nojiri03}.
The best agreement with the experiment was found by normalizing the
calculated intensity $S^{\mu}(\omega)$ in a way that its integral over
all frequencies at a fixed magnetic field equals unity. In all figures
presenting the calculated ESR spectrum such a normalized intensity is
visualized by the height of the peaks (in arbitrary units). For $B\|c$
(see Fig.~\ref{fig2}a) we obtain a finite spectral weight for the two
$S_z=-1$ triplet states $O_2$ and $O_1$. The $\delta\sim 90$~GHz
splitting between the $O_2$ and $O_1$ states is a consequence of the
finite $D_z^\prime=2.2~{\rm K}$. The level anti-crossing of the lowest
$O_2$ triplet with the ground state near critical field $B_c\sim 20$~T
is due to a finite value of intra-dimer DM interaction $D_x$, $D_y$,
since these are the only interactions that break the rotational
symmetry around the $z$-axis. These terms are also responsible for the
level anti-crossings at $B\sim 4.5$~T, $\nu=600$~GHz and $B\sim
2.5$~T, $\nu=800$~GHz.  Experimentally, only the later one is clearly
visible in Fig.~\ref{fig4}a of Ref.~\cite{nojiri03}. The overall
effect of finite values of $D_x$, $D_y$ can be clearly seen comparing
Figs.~\ref{fig2}a and \ref{fig3}a for which $D_x=D_y=0$. For $B\|c$
there is no level anti-crossing neither with the singlet ground state,
nor with excited singlets.

In principle, the $D_x^\prime$, $D_y^\prime$ terms could also
contribute to level anti-crossing of $O_1$ and (or) $O_2$ states with
the ground state.  However, a closer analytical calculation shows,
that these terms connect the ground state with a state, that consists
of a product of two triplet states, the one with $S_z=0$ and the other
with $S_z=1$, located on the two perpendicular dimers. This state 
is consequently orthogonal to $O_1$ and $O_2$ states. Therefore, one does
not expect level anticrossing in the first order in $D_x^\prime$,
$D_y^\prime$.
\begin{figure}[htb]
\includegraphics[angle=0,width=7.2cm,scale=1.0]{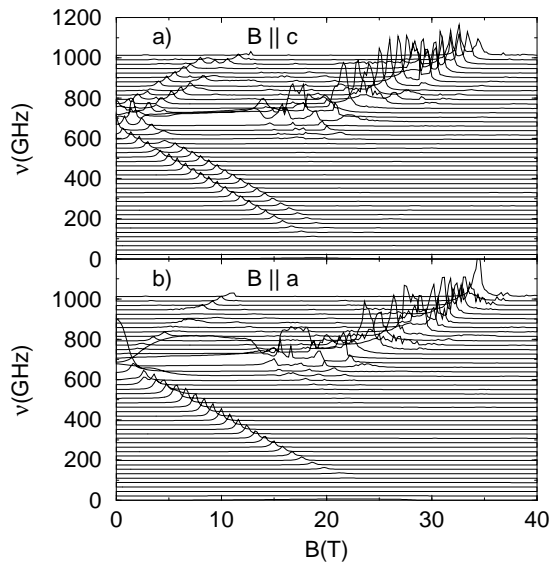}
\vspace{0.4cm}
\caption{Spin structure factor a) $S^x(\omega=2\pi\nu)$ for $B\|c$
and b) $S^z(\omega)$ for $B\|a$. Parameters of the model are:
$J=74~{\rm K}$, $J^\prime=0.62J$, ${\bf D}=(2.2~{\rm K},2.2~{\rm
K},3.7~{\rm K})$ and ${\bf D^\prime}=(0,0,2.2~{\rm K})$. Note that the
computed ESR spectrum is normalized as $\int\!\! d\omega\,
S^{\mu}(\omega)=1$ for any fixed $B$ ($\mu=x,z$).}
\label{fig2}
\end{figure}

The finite intensities of $O_1$ and $O_2$ triplets for $B\|c$ are a
consequence of a non-zero value of $D_z$ which requires a lower
crystal symmetry than the one observed with X-rays \cite{sparta01}.
In Fig.~\ref{fig4} we show the calculated ESR spectra for $D_z=0$.
The $O_1$ and $O_2$ triplet lines are not observed for $B\|c$ while in
$B\|a$ case, the lowest triplet excitations are clearly visible.
Finite values of $D_x$, $D_y$ or $D_x^\prime$, $D_y^\prime$ do not
induce these transitions in the lowest order. The reason is that they
mix $S_z=\pm 1$ states with the ground state. The non-zero $D_z$
term is therefore the only term within the given Hamiltonian
(\ref{Hamil}) that leads to $O_1$ and $O_2$ transitions for $B\|c$.

Comparing results of the model (\ref{Hamil}) with ``optimally'' chosen
parameters (see Fig.~\ref{fig2}) with the experiment in
Ref.~\cite{nojiri03} reveals a good agreement for the line positions,
and in some cases even matching of level anti-crossings with singlet
states. The main disagreement with the experiment is in line
intensities. While on the one hand ESR measurements of
Ref.~\cite{nojiri03} show that for $B\|c$ the $O_1$ line is nearly
$B$-field independent, $O_2$ line shows rather strong field
dependence: the intensity of the line increases with the applied
magnetic field. On the other hand, our numerical calculations in
Fig.~\ref{fig2}a show a nearly constant line intensities for both
$O_1$ and $O_2$ lines.
\begin{figure}[htb]
\includegraphics[angle=0,width=7.2cm,scale=1.0]{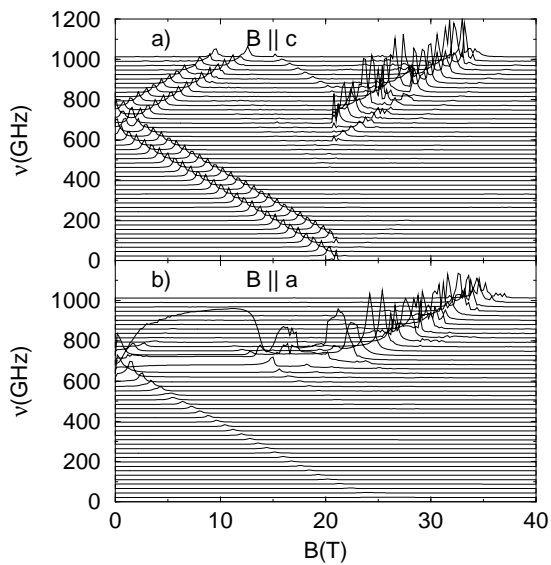}
\vspace{0.4cm}
\caption{The same as in Fig.~\ref{fig2} except 
${\bf D}=(0,0,3.7~{\rm K})$.}
\label{fig3}
\end{figure}
\begin{figure}[htb]
\includegraphics[angle=0,width=7.2cm,scale=1.0]{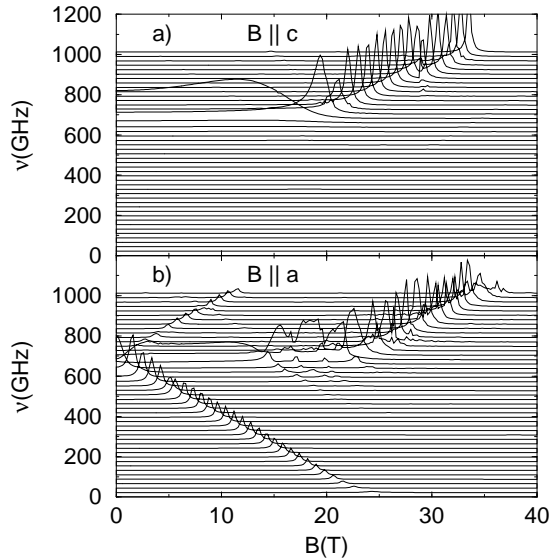}
\vspace{0.4cm}
\caption{The same as in Fig.~\ref{fig2} except
${\bf D}=(2.2~{\rm K},2.2~{\rm K},0)$.}
\label{fig4}
\end{figure}

We focus now on the case $B\|a$ which is shown in Fig.~\ref{fig2}b.
The $O_1$ and $O_2$ triplets are nearly degenerate. A level
anti-crossing with the ground state is well seen around $B\geq 20$~T.
The intensity of the $S_z=-1$ lines is varying non monotonously with
$B$. This non monotonous behavior could be due to a weak level
anti-crossing with localized singlets at $\nu=600$~GHz observed in
Fig.~\ref{fig1}b. Pronounced level anti-crossings are obtained in the
upper triplet branches with $S_z=1$. In contrast to the $B\|c$ case,
$D_z$ is not the only term that leads to transitions between the
ground state and the excited triplet states. In Figs.~\ref{fig3}b and
\ref{fig4}b we show $S^z(\omega)$ for $B\|a$ and ${\bf
D}=(0,0,3.7~{\rm K})$ and ${\bf D}=(2.2~{\rm K},2.2~{\rm K},0)$,
respectively. In both cases we see finite intensities of $O_1$ and
$O_2$ triplets, however ${\bf D}=(0,0,3.7~{\rm K})$ leads to a smaller
intensity than ${\bf D}=(2.2~{\rm K},2.2~{\rm K},0)$. Note that even
though intensities are presented in arbitrary units, the scaling of
intensities in all figures is identical to allow comparison.

\section{Conclusions}

In summary, the ESR spectra predicted by model (\ref{Hamil}) reproduce
several aspects of the experimental data obtained by Nojiri {\it et
al.} \cite{nojiri03} for SrCu$_{2}$(BO$_{3}$)$_{2}$. In particular,
for $B\|c$, the crystal symmetry breaking interaction $D_z$ is the
only term that leads to finite ESR intensities for $O_1$ and $O_2$
triplet states. We have tested other possible scenarios that could
provide finite ESR intensities for the low-lying triplet
excitations. One possibility is the introduction of an anisotropic
gyromagnetic $g$-tensor (Zeeman term in Eq.~(\ref{Hamil})) with a
different orientation for all 4 spins in the unit cell
\cite{kodama}. In order to get a finite ESR line for $B\|c$, the
external field coupled to the $g$-tensor would have to induce a
staggered field along the magnetic $z$-axis, such that each spin in a
dimer would feel different field orientation.  However, due to the
particular structure of the $g$-tensor, which is a consequence of the
buckling of the $ab$ planes in SrCu$_{2}$(BO$_{3}$)$_{2}$
\cite{kodama}, field $B\|c$ only induces staggered field component
along the $x$- and $y$-axis. A second possibility is the existence of
a small finite angle $\theta$ between the crystallographic $c$-axis
and the direction of the applied magnetic field ${\bf B}$ due to an
error in the orientation of the crystal. Taking into account that the
off-diagonal component of the $g$-tensor is at most of the order of
$g_s\sim 0.05g\sim 0.1$, we found almost no detectable signal for
$\theta<5^{\rm o}$.

The inclusion of the DM interaction $D_z$ provides the simplest way to
explain some qualitative aspects of the ESR experiments for
SrCu$_{2}$(BO$_{3}$)$_{2}$. This simple explanation has very important
experimental consequences. The existence of nonzero $D_z$ suggests that
the system should undergo a structural phase transition at low
temperatures that lowers the crystal symmetry. In the new phase, the
planes containing the $c$-axis and one dimer are no longer mirror
planes. In addition, we also expect a strong spin-lattice coupling
when the $O_1$ and $O_2$ triplet states get close to the singlet
ground state. Such a coupling could contribute to the stabilization of
the different plateaus that are observed in the magnetization
vs. field experiments \cite{kageyama99,onizuka00}.

\acknowledgements We wish to thank D. Ar\v con and A. Zorko for
fruitful discussions as well as S. Zvyagin and H. Nojiri for their
kind help with explaining the experimental setup.  J.~B., S. El
S. and I.~S. acknowledge the financial support of Slovene MESS under contract
P1-0044.  This work was in part sponsored by the US DOE under contract
W-7405-ENG-36.

\end{document}